\documentclass{article}
\usepackage{amsmath}
\usepackage{wrapfig}
\usepackage{pgfplots}
\usepackage{enumerate}
\usepackage{amssymb}
\usepackage{algorithm}
\input{Qcircuit}
\usepackage[noend]{algpseudocode}
\usepackage{amsmath}
\usepackage[justification=centering]{caption}
\numberwithin{equation}{section}
\usepackage{authblk}
\usepackage{hyperref}

\usepackage{amsthm}
 
\theoremstyle{definition}
\newtheorem{definition}{Definition}[section]

\usepackage{amsmath,amssymb}
\makeatletter
\newsavebox\myboxA
\newsavebox\myboxB
\newlength\mylenA

\newcommand*\xoverline[2][0.75]{%
    \sbox{\myboxA}{$\m@th#2$}%
    \setbox\myboxB\null
    \ht\myboxB=\ht\myboxA%
    \dp\myboxB=\dp\myboxA%
    \wd\myboxB=#1\wd\myboxA
    \sbox\myboxB{$\m@th\overline{\copy\myboxB}$}
    \setlength\mylenA{\the\wd\myboxA}
    \addtolength\mylenA{-\the\wd\myboxB}%
    \ifdim\wd\myboxB<\wd\myboxA%
       \rlap{\hskip 0.5\mylenA\usebox\myboxB}{\usebox\myboxA}%
    \else
        \hskip -0.5\mylenA\rlap{\usebox\myboxA}{\hskip 0.5\mylenA\usebox\myboxB}%
    \fi}
\makeatother

\title{A Quantum Algorithm for Finding Common Matches Between Databases with Reliable Behavior}

\author{Khaled El-Wazan \thanks{khaled\_elwazan@alex-sci.edu.eg}}

\affil{Department of Mathematics and Computer Science, Faculty of Science, Alexandria University, Egypt}

\date{}

\begin{document}
\maketitle

\begin{abstract}
Given $\kappa$ databases of unstructured entries, we propose a quantum algorithm to find the common entries between those databases. The proposed algorithm requires $\mathcal{O}(\kappa \sqrt{N})$ queries to find the common entries, where $N$ is the number of records for each database. The proposed algorithm constructs an oracle to mark common entries, and then uses a variation of amplitude amplification technique with reliable behavior to increase the success probability of finding them.
\end{abstract}

\section{Introduction}
\label{introduction}

Given  $\kappa$  databases with unstructured entries, it is required to find the joint entries between those databases.  Considering this problem in classical computers, an intuitive approach is to count similar entries from those databases and store them in a memory which keeps track of each entry and its number of occurrences, and then iterate  over this memory and observe when the number of occurrences of certain entries equal to $\kappa$. This procedure requires at most $\mathcal{O}(\kappa N)$ queries.

Quantum computers \cite{loyed,feynman,deutsch} are inherently probabilistic devices which promise to significantly accelerate certain types of computations compared to classical computers \cite{Bernstein-Vazirani}, by utilizing quantum phenomena like entanglement and superposition. Many quantum algorithms have emerged recently, for example, Deutch-Jousza algorithm \cite{Deutsch-jouza} that tests whether a given Boolean function  is a balanced Boolean function or a constant Boolean function, using only a single oracle call. P. Shor introduced a polynomial-time algorithm \cite{shor} to factorize a composite integer to its prime factors. L. Grover presented a quantum algorithm \cite{grover} to search for an entry in an unstructured list of entries in quadratic speed-up compared to classical computers.

In 1998, Burhman \textit{et al.} introduced an algorithm \cite{buhrman} that solves a problem similar to the common entries problem: given two remotely separated schedules of unknown free slots out of $N=2^n$ slots, find a common slot between those two schedules, in as minimum communication bits sent as possible. Burhman \textit{et al.}  algorithm requires $\mathcal{O}(\sqrt{N}\log_2 N)$ communication complexity and $\mathcal{O}(k\sqrt{N})$ query calls, with $k$ trails and error at most $2^{-k}$ \cite{buhrman}. Later in 2002, L. Grover proposed an algorithm \cite{grover-common} to solve scheduling problem with $\mathcal{O}(\sqrt{\epsilon N}\log_2{N})$ computation complexity.

In 2012, A. Tulsi proposed a quantum algorithm \cite{tulsi} to find a single common element between two sets in $\mathcal{O}(\sqrt{N})$ using an ancilla qubit to mark the common solution with phase-shift and applying amplitude amplification algorithm to increase the success probability of the desired result.

In 2013, Pang \textit{et al.} introduced a quantum algorithm \cite{pang} for set operations. In that literature, Pang \textit{et al.} provided a subroutine to find common intersected elements between two sets of size $2^n$ and $2^m$ elements in $\mathcal{O}(\sqrt{2^{m+n}/ C})$, where $C$ is the number of common entries, using a similar algorithm proposed in \cite{quant-tightbound-for-search}.

The aim of this paper is to propose an algorithm to find the common matches $M$ between given $\kappa$ databases each of $N$ entries. Each given database uses a black-box to identify its elements.   The proposed algorithm can find  a match among the common entries using a new oracle $U_\hbar$ which is constructed from the set of all given black-boxes $\hbar$. The new oracle $U_\hbar$ is then used along with amplitude amplification technique based on partial diffusion operator, to increase the success probability of finding the desired results. As well, the algorithm works  with probability of success at least $2/3$.

The paper is organized as follows: Section ~\ref{amplitude-amplification} depicts an amplitude amplification algorithm with reliable behavior used to solve the problem at hand. Section ~\ref{operator-construction} cover the construction of the oracle $U_\hbar$. Section ~\ref{proposed-algorithm} introduces the proposed algorithm. Section ~\ref{analysis} analyzes the proposed algorithm. Section \ref{comparison} compares the proposed algorithm to other literature, followed by a conclusion in Section ~\ref{conclusion}.

\section{Amplitude Amplification}
\label{amplitude-amplification}

Consider having a list $L$ of $N=2^n$ of unstructured entries, which has an oracle $U_f$ that is being used to access those entries. Each entry $i\in L= \{0,1,..., N-1\}$ in the list $L$ is mapped to either $0$ or $1$ according to any certain property satisfied by $i$ in $L$, \textit{i.e.} $f:L\rightarrow\{0,1\}$. The amplitude amplification problem is stated as follows: find the entry $i\in L$ such that $f(i)=1$.

In 1996, L. Grover proposed a unique approach to solve this typical problem  with quadratic speed-up compared to classical algorithms \cite{grover}.
The algorithm Grover proposed takes advantage of quantum parallelism to solve this problem by preparing a perfect superposition of all the possible $N$ entries corresponding to the list $L$, after that it starts marking the solution using phase shift of $-1$ using the oracle $U_f$, followed by amplifying the amplitude of the solution using inversion about the mean operator. It was shown in \cite{grover,grover-younes} that the algorithm requires $\pi/4\sqrt{N}$ iteration to optimally \cite{grover-optimal} find a solution to the search problem with high probability, assuming there is only one solution $i\in L$ that satisfies the oracle $U_f$.

Boyer \textit{et al.} later generalized Grover's quantum search algorithm  to fit the purpose of finding multiple solutions $M$ to the oracle $U_f$, \textit{i.e.} $\forall p$, for which $1\leq p\leq M\leq 3N/4$, $f(i_p)=1$, to require a number of $\pi/4\sqrt{N/M}$ iterations of the algorithm \cite{quant-tightbound-for-search}. For the case of unknown number of solutions $M$ to the oracle, an algorithm \cite{quant-count} was proposed to find such number $M$. However, the generalized quantum search algorithm has shown to exponentially fail in the case of $M>3N/4$ \cite{quant-tightbound-for-search,grover-younes}.

Younes \textit{et al.} introduced a variation of the generalized quantum search algorithm \cite{younes} with reliable behavior in case of multiple solutions to the oracle $U_f$, \textit{i.e.} $1\leq M\leq N$, and requires $\mathcal{O}(\sqrt{N/M})$ oracle calls.

In the case of known multiple solutions $M$ for a list $L$ of size $N=2^n$, Younes \textit{et al.} algorithm is outlined as follows:
\begin{figure}[H]
\begin{align*}
 \Qcircuit @C=1em @R=.7em {
  \lstick{\ket{0}} & /^n \qw & \gate{H^{\otimes n}} & \multigate{1}{U_f} & \qw & \multigate{1}{Y} & /^n \qw 	& \meter &  \qw  \\
  \lstick{\ket{0}} & \qw     & \qw   & \ghost{U_f}   & \qw & \ghost{Y}     & \qw  \gategroup{1}{4}{2}{6}{.7em}{_\}} \\
  \\
  & & & & & \lstick{\mathcal{O}(\sqrt{N/M})} & &
 }
\end{align*}
\caption{Quantum circuit for the quantum search algorithm \cite{younes}.\label{quantum-search-circuit}}
\end{figure}
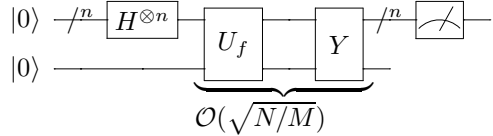

\begin{enumerate}

\item Prepare a quantum register with $n+1$ qubits in a uniform superposition:

\begin{equation}
|\varphi\rangle = \frac{1}{\sqrt{N}}\sum_{i=0}^{N-1}|i\rangle \otimes |0\rangle.
\end{equation}

\item Iterate the algorithm for ${\pi}/{(2\sqrt{2})}\sqrt{{N}/{M}}$ times by applying the partial diffusion operator $Y$ on the state $U_f|\varphi\rangle$ in each iteration, such that it performs the inversion about the mean on a subspace of the system, where

\begin{equation}
Y=(H^{\otimes n}\otimes I)(2|0\rangle\langle 0|-I_{n+1})(H^{\otimes n}\otimes I).
\end{equation}

At any iteration $q\geq 2$, the system can be described as follows \cite{younes}:
\begin{equation}
|\varphi^q\rangle=a_q\sum_{i=0}^{{N-1}}\strut^{\prime\prime}\big(|i\rangle\otimes|0\rangle\big)+b_q\sum_{i=0}^{{N-1}}\strut^{\prime}\big(|i\rangle\otimes|0\rangle\big)+c_q\sum_{i=0}^{{N-1}}\strut^{\prime}\big(|i\rangle\otimes|1\rangle\big),
\end{equation} 
where the amplitudes $a_q,~b_q$ and $c_q$ are recursively defined as follows:
\begin{align}
a_q&=2\langle\alpha_q\rangle-a_{q-1}, \quad
b_q=2\langle\alpha_q\rangle-c_{q-1}, \quad
c_q=-b_{q-1},
\end{align}
and
\begin{equation}
\langle\alpha_q\rangle=\Big(\big(1-\frac{M}{N}\big)a_{q-1}+\big(\frac{M}{N}\big) c_{q-1}\Big).
\end{equation}

\end{enumerate}

For this algorithm, the success probability is as follows \cite{younes}:

\begin{equation}
P_s=\big(1-\cos\big(\theta\big)\big)\Big(\frac{\sin^2\big(\big(q+1\big)\theta\big)}{\sin^2\big(\theta\big)}+\frac{\sin^2\big(q\theta\big)}{\sin^2\big(\theta\big)}\Big),
\end{equation}
where $\cos\big(\theta\big)=1-M/N$, $0< \theta\leq\pi/2$, and the required number of iterations $q$ is given by \cite{younes}:

\begin{equation}
q=\Bigl\lfloor\frac{\pi}{2\theta}\Bigr\rfloor \leq\frac{\pi}{2\sqrt{2}}\sqrt{\frac{N}{M}},
\label{q-younes}
\end{equation}
where $\lfloor~ \rfloor$ is the floor operation.

Although Younes \textit{et al.} variation of quantum search algorithm runs slower compared to Grover's algorithm by $\sqrt{2}$ for small $M/N$, but Younes \textit{et al.} algorithm is more reliable with high probability than generalized Grover search algorithm for multiple matches $M$ \cite{younes} such that $1\leq M\leq N$.

\section{Constructing the Oracle $U_\hbar$}
\label{operator-construction}
%
In this section, the given set of oracles $\hbar$ 
will be utilized to construct the oracle $U_\hbar$ which will be used for finding the common solutions $M$ between the oracles in the set $\hbar$, assuming that all the given oracles are of $N=2^n$ unstructured entries, given that $n$ is the number of inputs to all of the given oracles. For the sake of simplification, we will provide a simple illustration for the oracle $U_\hbar$ assuming that the size of the set $\hbar$ is only $\kappa=2$ oracles, and after that we will propose the generalized form of the oracle $U_\hbar$ for multiple oracles $\kappa\geq 2$.

\begin{definition}

Let's assume having a Boolean function $f$ that maps a vector of size $n$ to either $0$ or $1$, \textit{i.e.} $f:\{0,1\}^n\rightarrow \{0,1\}$. An oracle $U_f$ is defined to perform such mapping. We say that $U_f$ is an operator on $n+t+q+1$ qubits, taking the control $0\rightarrow n-1$ qubits and targets the qubit with the index $n+t$; this configuration will be denoted as $^{0\rightarrow n-1}_{n+t}U_f$. Such defined oracle can be illustrated as follows:

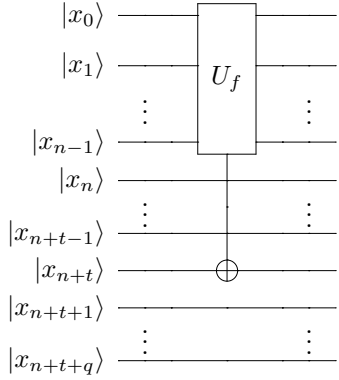
\begin{figure}[H]
\begin{align*}
\Qcircuit @C=1em @R=1em { 
 \lstick{\ket{x_0}} & \qw  & \qw  & \multigate{3}{U_f}& \qw  & \qw  & \qw \\
 \lstick{\ket{x_1}} & \qw  & \qw  & \ghost{U_f}& \qw  & \qw   & \qw\\ 
  &  \vdots &  &&& \vdots\\ 
 \lstick{\ket{x_{n-1}}} & \qw  & \qw  & \ghost{U_f}& \qw  & \qw  & \qw \\ 
 \lstick{\ket{x_n}} & \qw  & \qw  & \qwx \qw  & \qw  & \qw   & \qw\\ 
 &  \vdots & &\qwx && \vdots\\ 
 \lstick{\ket{x_{n+t-1}}} & \qw  & \qw  & \qwx  \qw   & \qw  & \qw  & \qw \\ 
 \lstick{\ket{x_{n+t}}}& \qw  & \qw  & \targ \qwx &\qw  & \qw  & \qw  \\ 
 \lstick{\ket{x_{n+t+1}}}& \qw  & \qw  & \qw  & \qw  & \qw   & \qw \\ 
 &  \vdots & &&& \vdots\\ 
 \lstick{\ket{x_{n+t+q}}}& \qw  & \qw  & \qw  & \qw  & \qw  & \qw \\ 
}
\end{align*}
\caption{ A quantum circuit representing the oracle $\strut^{0\rightarrow n-1}_{n+t}U_f$.}
\end{figure}

\end{definition}

For the problem of finding common entries $M$ between $\kappa$ oracles, the problem statement can be defined as follows:
\begin{definition}
\label{DB-with-same-size-pd}
Consider having a set $\mathcal{Z}$ of $\kappa\geq 2$ lists, $\mathcal{Z}=\{L_0,\cdots,L_{\kappa-1}\}$. Each list $L_j\in \mathcal{Z}$ is of $N=2^n$ unstructured entries, which has an oracle $U_j$ that is being used to access those entries in $L_j$. Each entry $i\in L_j=\{0,1,\cdots,N-1\}$ in the list $L_j$ is mapped to either $0$ or $1$ according to any certain property satisfied by $i$ in $L_j$, \textit{i.e.} $f_j:L_j\rightarrow\{0,1\}$. The common elements problem is stated as follows: find the entry $i\in L_j $  such that $ \forall L_j\in \mathcal{Z},~ f_j(i)=1$.
\end{definition}

\subsection{Constructing the Oracle $U_\hbar$ for Two Databases}
\label{Uhbar-for-same-2-databases}

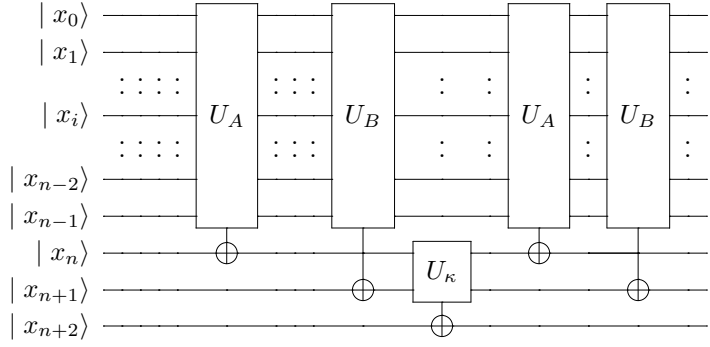
\begin{figure}[H]
\begin{align*}
\Qcircuit @C=0.7em @R=.5em  {
\lstick{\mid x_0 \rangle}& \qw & \qw & \qw & \qw & \multigate{8}{U_A}   & \qw & \qw & \qw & \multigate{8}{U_B} & \qw & \qw & \multigate{8}{U_A} & \qw & \multigate{8}{U_B} & \qw &\qw\\
\lstick{\mid x_1 \rangle}& \qw & \qw & \qw & \qw & \ghost{U_A}  & \qw & \qw & \qw & \ghost{U_B} & \qw & \qw & \ghost{U_A}& \qw & \ghost{U_B}& \qw  & \qw\\
& . & . & . & . &  & .  & . & . &  & .  & . &  & . && .\\
& . & . & . & . &  & .  & . & . &  & .  & . &  & . && .\\
\lstick{\mid x_i \rangle} & \qw & \qw & \qw & \qw & \ghost{U_A} & \qw & \qw & \qw & \ghost{U_B} & \qw & \qw & \ghost{U_A}& \qw & \ghost{U_B}& \qw  & \qw\\
& . & . & . & . &  & .  & . & . &  &  . & . &   & .&& .\\
& . & . & . & . &  & .  & . & . &  &   .& . &   & .&& .\\
\lstick{\mid x_{n-2} \rangle}& \qw & \qw & \qw & \qw & \ghost{U_A}  & \qw & \qw & \qw & \ghost{U_B} & \qw & \qw & \ghost{U_A}& \qw & \ghost{U_B}& \qw  & \qw \\
\lstick{\mid x_{n-1} \rangle}& \qw & \qw & \qw & \qw & \ghost{U_A}  & \qw & \qw & \qw & \ghost{U_B} & \qw & \qw & \ghost{U_A}& \qw & \ghost{U_B}& \qw & \qw \\
\lstick{\mid x_n \rangle}& \qw & \qw & \qw & \qw &  \targ{-1} \qwx & \qw  & \qw & \qw  & \qw \qwx & \multigate{1}{U_\kappa} & \qw & \targ \qwx & \qw &\qw \qw \qwx & \qw & \qw \\
\lstick{\mid x_{n+1} \rangle}& \qw & \qw & \qw & \qw &  \qw & \qw  & \qw & \qw & \targ \qwx & \ghost{U_\kappa} & \qw  & \qw & \qw &\targ \qwx &\qw& \qw  \\
\lstick{\mid x_{n+2} \rangle}& \qw & \qw & \qw & \qw &  \qw & \qw  & \qw & \qw & \qw & \targ \qwx & \qw  & \qw & \qw  & \qw& \qw & \qw \\
}
\end{align*}
\caption{ A quantum circuit for the proposed oracle $U_\hbar$ for $\kappa=2$ functions.\label{fig:proposed-oracle-circuit-2fns}}
\end{figure}

Given that $\kappa=2$ oracles, $U_A$ and $U_B$, which map the elements of black-box functions $f_A$ and $f_B$ of $n$ input to either $0$ or $1$, it is required to find the common solutions $M$ between them. It will be required to reserve $3$ auxiliary qubits.
An illustration of this circuit is shown in Figure ~\ref{fig:proposed-oracle-circuit-2fns}.
 
A quantum circuit for the oracle $U_\hbar$ can be constructed as follows: 
\begin{equation}
U_\hbar =\strut^{0\rightarrow n-1} _{n+1}U_B \times \strut^{0\rightarrow n-1}_{n}U_A \times \strut^{n\rightarrow n+1} _{n+2}U_\kappa\times \strut^{0\rightarrow n-1}_{n+1}U_B \times \strut^{0\rightarrow n-1}_{n}U_A,
\end{equation}
 where  the operator $U_\kappa$ represents the function $f_\kappa(x)$:
 
\begin{equation}
f_\kappa(x)=f_A(x)\cdot f_B(x),\label{fkappa-2}
\end{equation}
 such that $\cdot$ is the AND logic operation, and $x\in\{0,1\}^n$.

To clarify the effect of the proposed oracle $U_\hbar$, let's analyze that effect on a uniform superposition as follows:

\begin{enumerate}
\item \textit{Register Preparation.} Prepare a quantum register of size $n+3$ qubits in the state $\vert 0\rangle$, where the last $3$ qubits will be utilized as extra space to compute the oracles $U_A$, $U_B$ and the common solutions between them:

\begin{equation}
\vert\varphi_0\rangle=\vert 0\rangle^{\otimes n}\otimes \vert 0\rangle^{\otimes 3}.
\end{equation}

\item \textit{Register Initialization.} Apply Hadamard gates on the first $n$ qubits to get a uniform superposition of all the possible $N=2^n$ states:

\begin{align}
\vert\varphi_1\rangle&=H^{\otimes n}\vert\varphi_0\rangle \nonumber \\
&=H^{\otimes n}\vert 0\rangle^{\otimes n} \otimes \vert 0\rangle^{\otimes 3}\nonumber \\
&=\frac{1}{\sqrt{N}}\sum_{i=0}^{N-1}\vert i\rangle \otimes \vert 0\rangle^{\otimes 3}.
\end{align}

\item \textit{Applying the Oracle $U_A$.} Apply the oracle $U_A$ on the register to mark all its possible solutions 
in the first extra qubit, where non-solutions will be marked with $\vert 0\rangle$ and the solutions will be marked with $\vert 1\rangle$: 

\begin{align}
\vert \varphi_2\rangle&=\strut^{0\rightarrow n-1} _{n}U_A \vert\varphi_1\rangle \nonumber \\
&=\frac{1}{\sqrt{N}}\sum_{i=0}^{N-1}\vert i\rangle \otimes \vert f_A(i)\rangle \otimes \vert 0\rangle^{\otimes 2}.
\end{align}

\item \textit{Applying the Oracle $U_B$.} Apply the oracle $U_B$ on the register to mark all its possible solutions 
in the second extra qubit, where the non-solution states will be marked with $\vert 0\rangle$ and the solution states will be marked with $\vert 1\rangle$:

\begin{align}
\vert \varphi_3\rangle&=\strut^{0\rightarrow n-1} _{n+1}U_B \vert\varphi_2\rangle \nonumber \\
&=\frac{1}{\sqrt{N}}\sum_{i=0}^{N-1}\vert i\rangle \otimes \vert f_A(i)\rangle \otimes \vert f_B(i)\rangle \otimes \vert 0\rangle.
\end{align}

\item \textit{Applying the Operator $U_\kappa$.} Apply the operator $U_\kappa$ on the register to mark all possible common solutions between the oracles $U_A$ and $U_B$ in the third extra qubit, where non-common solutions will be marked with $\vert 0\rangle$ and the common solutions will be marked with $\vert 1\rangle$: 

\begin{align}
\ket{\varphi_4}&=\strut^{n\rightarrow n+1} _{n+2}U_\kappa \vert\varphi_3\rangle \nonumber \\
&=\frac{1}{\sqrt{N}}\sum_{i=0}^{N-1}\vert i\rangle \otimes \vert f_A(i)\rangle \otimes \vert f_B(i)\rangle \otimes \vert f_\kappa(i)\rangle,
\end{align}
where $f_\kappa(i)$ is defined as in Equation ~\eqref{fkappa-2}.

\item \textit{Applying $U_BU_A$.} Apply both the oracles $U_BU_A$ to remove any entanglement between the solutions of both oracles from the first and the second extra qubits, and reset them to their initial state $\ket{0}^{\otimes 2}$:

\begin{align}
\ket{\varphi_5}&=\strut^{0\rightarrow n-1} _{n+1}U_B \times \strut^{0\rightarrow n-1} _{n}U_A  \vert\varphi_4\rangle \nonumber \\
&=\frac{1}{\sqrt{N}}\sum_{i=0}^{N-1}\vert i\rangle \otimes \vert 0\rangle^{\otimes 2} \otimes \vert f_\kappa(i)\rangle.
\label{marking-common-soln}
\end{align}

Ignoring the reset extra qubits, the state $\ket{\varphi_5}$ can be rewritten as follows:
\begin{equation}
\ket{\varphi_5}=\frac{1}{\sqrt{N}}\sum_{i=0}^{N-1}\strut^{\prime\prime}(\vert i\rangle \otimes \vert 0\rangle)+ \frac{1}{\sqrt{N}}\sum_{i=0}^{N-1}\strut^{\prime}(\vert i\rangle\otimes \vert 1\rangle),
\label{marking-common-soln-1}
\end{equation}
where $\sum^{\prime\prime}$ are all the possible uncommon solutions between the oracles $U_A$ and $U_B$ marked with $\vert 0\rangle$, and $\sum^\prime$ are all the possible common solutions between those oracles marked with $\vert 1\rangle$.

\end{enumerate}

The main reason behind applying each oracle for the second time on its target qubit at each call of $U_\hbar$, is that the solutions of that specific oracle are still entangled with their target qubit. Discarding that qubit at the stage of amplifying the common solutions will drastically affect the desired outcome of the algorithm \cite{monogamy1}. So to get rid of this  entanglement, applying each oracle on its respective target qubit is necessary to remove such correlation and maintain a valid result.

\subsection{The Oracle $U_\hbar$ for more than Two Databases}
\label{Uhbar-for-same-k-databases}
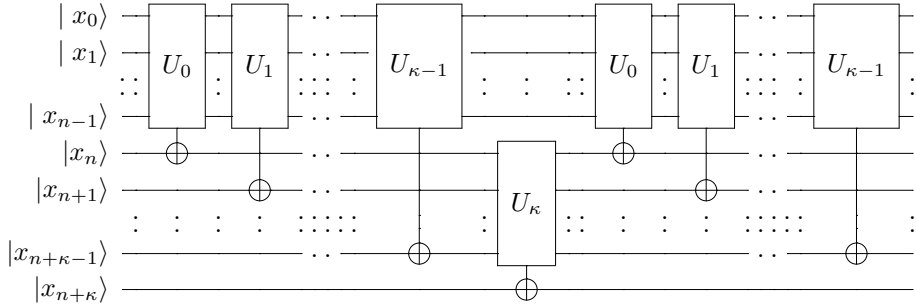
\begin{figure}[H]
\begin{align*}
\Qcircuit @C=0.5em @R=.5em  {
\lstick{\mid x_0 \rangle}& \qw& \multigate{4}{U_0} & \qw & \multigate{4}{U_1}& \qw & . & . & & \qw & \multigate{4}{U_{\kappa-1}} &  \qw&\qw& \qw&\qw &\multigate{4}{U_0} & \qw & \multigate{4}{U_1}& \qw & . & . & & \qw & \multigate{4}{U_{\kappa-1}}&\qw \\
\lstick{\mid x_1 \rangle}& \qw& \ghost{U_0} & \qw & \ghost{U_1}& \qw & . & . & & \qw & \ghost{U_\kappa-1} &  \qw&\qw& \qw&\qw &\ghost{U_0} & \qw & \ghost{U_1}& \qw & . & . & & \qw & \ghost{U_{\kappa-1}}&\qw \\
.& .&  & . &  & . & . & . & & . &   &  . & . & . & . &  & . &  & . & . & . & & . &  &. \\
.& .&  & . &  & . & . & . & & . &   &  . & . & . & . &  & . &  & . & . & . & & . &  &. \\
\lstick{\mid x_{n-1} \rangle}& \qw& \ghost{U_0} & \qw & \ghost{U_1}& \qw & . & . & & \qw & \ghost{U_{\kappa-1}} &  \qw&\qw& \qw&\qw &\ghost{U_0} & \qw & \ghost{U_1}& \qw & . & . & & \qw & \ghost{U_{\kappa-1}}&\qw \\
\lstick{\ket{x_n}} &\qw & \targ \qwx & \qw & \qw \qwx & \qw & . & . & & \qw & \qw\qwx & \qw & \multigate{4}{U_\kappa} & \qw & \qw  & \targ \qwx & \qw & \qw \qwx & \qw & . & . & & \qw & \qw\qwx & \qw & \\
\lstick{\ket{x_{n+1}}} &\qw & \qw & \qw & \targ \qwx & \qw & . & . & & \qw & \qw\qwx & \qw & \ghost{U_\kappa} & \qw & \qw  & \qw & \qw&\targ \qwx  & \qw & . & . & & \qw & \qw\qwx & \qw & \\
  & . & . & . & . & . & . & . & . & . & \qwx & . &  & . & . & . & . & . & . & . & . & .& . & \qwx & . \\
    & . & . & . & . & . & . & . & . & . & \qwx & . &  & . & . & . & . & . & . & . & . & .& . & \qwx & . \\
\lstick{\ket{x_{n+\kappa-1}}} &\qw & \qw & \qw & \qw & \qw & . & . & & \qw & \targ\qwx & \qw & \ghost{U_\kappa} & \qw & \qw  & \qw & \qw&\qw  & \qw & . & . & & \qw & \targ\qwx & \qw & \\
\lstick{\ket{x_{n+\kappa}}} &\qw & \qw & \qw&\qw & \qw & \qw&\qw & \qw & \qw&\qw & \qw & \targ\qwx  &\qw & \qw & \qw&\qw & \qw & \qw&\qw & \qw & \qw&\qw & \qw & \qw &
}
\end{align*}
\caption{A quantum circuit for the proposed oracle $U_\hbar$ for $\kappa$ functions.}
\label{fig:proposed-oracle-circuit-kfns}
\end{figure}

Given that $\kappa\geq 2$ oracles of $n$ input qubits and $\kappa+1$ auxiliary qubits, we illustrate the circuit of the oracle $U_\hbar$ in Figure ~\ref{fig:proposed-oracle-circuit-kfns}.

The oracle $U_\hbar$ can be generally defined as follows:
\begin{equation}
U_\hbar=\prod_{j=0}^{\kappa-1}\strut^{0\rightarrow n-1} _{n+j}U_j  \times  \strut^{n\rightarrow n+\kappa-1} _{n+\kappa}U_\kappa \times \prod_{j=0}^{\kappa-1}\strut^{0\rightarrow n-1} _{n+j} U_j,
\end{equation}
where $U_\kappa$ represents the function $f_\kappa(x)$ such that
\begin{equation}
f_\kappa(x) = \bigwedge_{j=0}^{\kappa-1}f_j(x),
\end{equation}
and $\bigwedge$ represents the AND logic operation.

The general system in a uniform superposition for $\kappa\geq 2$ after a single iteration, can be generally described as follows:
\begin{equation}
\vert \varphi\rangle=\frac{1}{\sqrt{N}}\sum_{i=0}^{N-1}\vert i\rangle \otimes \vert 0\rangle^{\otimes \kappa} \otimes \vert f_\kappa(i)\rangle.
\end{equation}

\section{The Proposed Algorithm}
\label{proposed-algorithm}
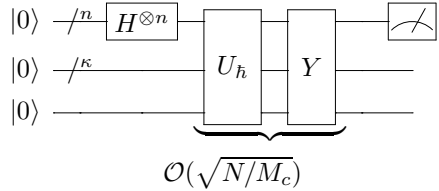
\begin{figure}[H]
\begin{align*}
\Qcircuit @C=1em @R=.7em {
  \lstick{\ket{0}} & /^n \qw & \gate{H^{\otimes n}} & \multigate{2}{U_{\hbar}} & \multigate{2}{Y}	 & \qw  & \meter \\
  \lstick{\ket{0}}   & /^\kappa  \qw  & \qw &  \ghost{U_\hbar}  &\ghost{Y} & \qw &\qw\\
   \lstick{\ket{0}}   & \qw  & \qw   &  \ghost{U_\hbar} & \ghost{Y}   & \qw & \qw   \gategroup{1}{4}{3}{5}{.7em}{_\}} & 
   \\
   &&&\dstick{\mathcal{O}(\sqrt{N/M_c})}& 
 }
\end{align*}
\caption{Quantum circuit for the proposed algorithm.
\label{proposed-algorithm-circuit}}
\end{figure}

In this section, we will propose the algorithm to find the common solutions $M_c$ such that $1\leq M_c\leq N$, among $\kappa$ oracles, based on Younes \textit{et al.} amplitude amplification algorithm. An illustration of the circuit is shown in Figure ~\ref{proposed-algorithm-circuit}.

The algorithm is carried quantum mechanically as follows:

\begin{algorithm}[H]
\label{1st-algo}
\begin{algorithmic}[1]
\State Construct the oracle $U_\hbar$.
\State Set the quantum register to $\ket{0}^{\otimes n}$ and the extra $\kappa+1$ qubits to $\ket{0}$.
\State Apply the Hadamard gates to the first $n$ qubits to create the uniform superposition $\frac{1}{\sqrt{N}}\sum_{i=0}^{N-1}{\ket{i}}\otimes \ket{0}^{\otimes \kappa+1}$.
\State \iterate Iterate over the following $q_c=\frac{\pi}{2\sqrt{2}}\sqrt{\frac{N}{M_c}}$ steps:
\begin{enumerate}
\item Apply the oracle $U_\hbar$.
\item Apply the diffusion operator $Y$.

\end{enumerate}
\State Measure the output.
\end{algorithmic}
\caption*{The Proposed Algorithm.}
\end{algorithm}

\section{Analysis of the Proposed Algorithm}
\label{analysis}
In this section, we will discuss the behavior of the proposed algorithm with respect to all possible scenarios for any given databases.

\subsection{In Case of Known Number of Common Matches between Databases}
Given that $\kappa\geq 2$ oracles, a single call to the oracle $U_\hbar$ will execute each given oracle $U_j$ exactly $2$ times. After the amplitude amplification of the desired common solutions, the total number of oracle calls $q_t$ for all given oracles can be expressed as the following sum:
\begin{align}
q_t&=\sum_{p=0}^{ q_c-1}2\times \kappa\nonumber \\
&=2\times \kappa\times \frac{\pi}{2\sqrt{2}}\sqrt{N/M_c}\nonumber\\
&= \kappa\times \frac{\pi}{\sqrt{2}}\sqrt{N/M_c} 
\end{align}

So, for any given $\kappa$ oracles with the same size, the number of oracle call to solve the common matches problem is $\mathcal{O}(\kappa\sqrt{N/M_c})$.

\subsection{In Case of Unknown Matches $M$}
An algorithm for estimating the number of matches was presented in \cite{quant-count}, known as \textit{quantum counting}. The proposed oracle $U_\hbar$ can be used with the quantum counting algorithm to estimate the number of matches $M$, before executing the proposed algorithm.

In \cite{younes}, another algorithm was presented by Younes \textit{et al.} to search for a match in a database, with unknown number of matches $M$ such that $1\leq M\leq N$. This algorithm can be combined with the proposed oracle $U_\hbar$ to find a common match, when the number of matches is not known in advance.

\section{Comparison with Other literature}
\label{comparison}

In 2012, Tulsi proposed an algorithm \cite{tulsi} that given two oracles that can identify the elements of two sets with the same size, the goal is to find a common element between those two sets. The success of finding that single element is further enhanced using a variation Tulsi introduced of Grover's amplitude amplification algorithm, with some restrictive conditions.

\subsection{Single Common Solution Amplification}
In the case of a single common solution between $\kappa=2$ oracles, Tulsi's algorithm is found to be optimal with restrictions, and requires $\mathcal{O}(\sqrt{N})$ oracle calls. However, the proposed algorithm requires the same oracle calls $\mathcal{O}(\sqrt{N})$ but with no restrictive conditions. 
In the case of single common solution when $\kappa> 2$ oracles which was not covered by Tulsi \cite{tulsi}, the proposed algorithm is found to require $\mathcal{O}(\kappa\sqrt{N})$ oracle calls.

\subsection{Multiple Common Solutions Amplification}
In the case of multiple common solutions between $\kappa=2$ oracles, the expected oracle calls of the proposed algorithm is $\mathcal{O}(\sqrt{N/M_c})$, when $M_c$ is $1\leq M_c\leq N$. Tulsi's algorithm can be used to cover the case of multiple solutions when $\kappa=2$, but the problem becomes exponentially harder when $M_c> 3N/4$ \cite{quant-tightbound-for-search,grover-younes}.
In the case of multiple common solutions between $\kappa\geq2$ oracles, this case is not covered by Tulsi ~\cite{tulsi}, however, the proposed algorithm requires $\mathcal{O}(\kappa\sqrt{N/M_c})$ oracle calls.

\section{Conclusion}
\label{conclusion}

In this paper, we proposed an algorithm to find the common entries $M$ between $\kappa$ databases. 
Each database uses an oracle to access its entries.  It is shown that the given oracles is used to construct another oracle that exhibits the behavior of finding only the common entries between those databases. The constructed oracle is used to mark the common entries with entanglement, then an amplitude amplification algorithm is applied to increase the success probability of finding the common entries.

It is found that when the given $\kappa$ databases are of the same size, it will require $\mathcal{O}(\kappa\sqrt{N/M_c})$ oracle calls.
As well, It is found that the performance of the proposed algorithm is more reliable in the case of multiple matches and quadratically faster than other literature solving this problem, and handles the general case of multiple databases with similar sizes.
The proposed oracle can be extended using \cite{quant-count} to count the number of common entries between any given oracles, or find a match as in \cite{quant-tightbound-for-search,younes}, when the number of common entries $M$ is unknown.

\bibliographystyle{IEEEtran}
\bibliography{references}

\end{document}